# Thermoelectric and galvanomagnetic properties of bismuth chalcogenide nanostructured hetero-epitaxial films

L N Lukyanova[1], Yu A Boikov[1], V A Danilov[1], O A Usov[1],

M P Volkov[1, 2], and V A Kutasov[1]

[1]Ioffe Physical-Technical Institute, Russian Academy of Sciences, Polytekhnicheskaya 26, 194021 St.-Petersburg, Russia

[2]International Laboratory of High Magnetic Fields and Low Temperatures, 53-421, Wroclaw, Poland

E-mail: lidia.lukyanova@mail.ioffe.ru

**Abstract**

Hot wall technique was used to grow block single crystal films of $Bi_2Te_3$ and solid solutions of $Bi_{0.5}Sb_{1.5}Te_3$ on mica (muscovite) substrates. X-ray diffraction studies demonstrated that the crystalline c-axis in the films was normal to the substrate plane. Seebeck coefficient, electrical conductivity and magnetoresistivity tensor components were measured at various orientations of magnetic and electric fields in the temperature interval 77-300 K and magnetic field up to 14 T. Scattering mechanism of charge carriers in the films were studied using temperature dependences of the degeneracy parameter and the Seebeck coefficient in terms of a many-valley model of energy spectrum. Obtained results have shown that the effective scattering parameter is considerably differed from the value specific for an acoustic scattering of charge carriers in the weakly degenerate films due to an additional scattering of charge carriers on interface and interctystallite boundaries. These features of charge carrier scattering are supposed to affect electronic transport in the films and enhance figure of merit.

## 1. Introduction

Thermoelectric epitaxial films based on bismuth and antimony chalcogenides are known to provide an enhanced thermoelectric figure of merit ZT as compared with that of the bulk thermoelectrics due to selective phonon scattering on interface and intercrystallite grain boundaries [1-6].

For development and optimization of a technology of nanostructured epitaxial films based on high efficiency bulk thermoelectrics [7-11], the charge carrier scattering mechanisms were investigated. Features of the scattering mechanisms in films under consideration may be revealed from the analysis of galvanomagnetic properties and the Seebeck coefficient as it has been carried out for bulk thermoelectrics of the same compositions [12].

These results are considered to be useful to obtain films with optimal thermoelectric power factor that together with the reduction of lattice thermal conductivity induced by selective phonon scattering in low-dimensional films provide an enhancement of thermoelectric figure of merit of nanostructured chalcogenides films.

## 2. Samples for studies and structural investigations

Hot wall technique [13] was used to grow 50-500 nm thick epitaxial films of $Bi_{2-x}Sb_xTe_3$ (x=0; 1.5) on (000l) mica (muscovite) substrates. An endemic feature of the hot wall technique is abnormally high substrate temperature ($T_S$) during a film growth. High $T_S$ promote enhancement of mobility of the particles (atoms, molecules and so on) adsorbed at a substrate surface. $T_S$ during thermoelectric layer formation was diminished just on 150 C as compared with temperature of the sublimating $Bi_{2-x}Sb_xTe_3$ powder. To eliminate vapor phase condensation at the growth chamber walls, temperature of the last was about $T_S$+50C. An effective growth rate of the thermoelectric layer was in the range 1-3 Å/s.

As follows from Atomic Force Microscopy (AFM) images of the substrate surface at initial stages of the layer growth, rectangular stable nuclei of the $Bi_{2-x}Sb_xTe_3$ were well laterally preferentially oriented, see insert in figure 1. Lateral size of the stable nuclei just before formation of continues thermoelectric layer is in the range 1-15 μm.

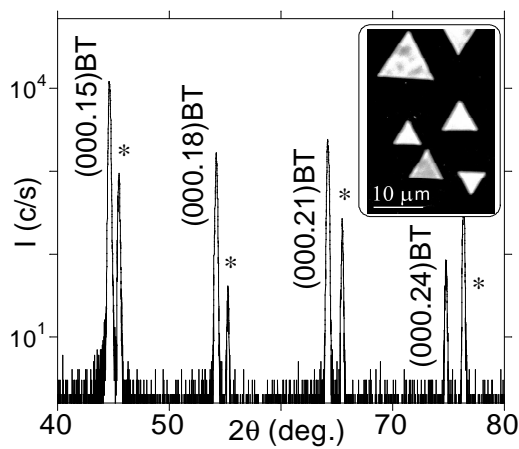

**Figure 1.** X-ray (CuK$_{α1}$, ω/2θ-) scan of the (300 nm) $Bi_2Te_3$ film grown at a mica substrate. *- peaks from the substrate. Triangular stable nuclei of the $Bi_2Te_3$ are clear resolved at the surface of the mica during initial stage of the thermoelectric layer growth, see insert. The nuclei are laterally preferentially oriented.

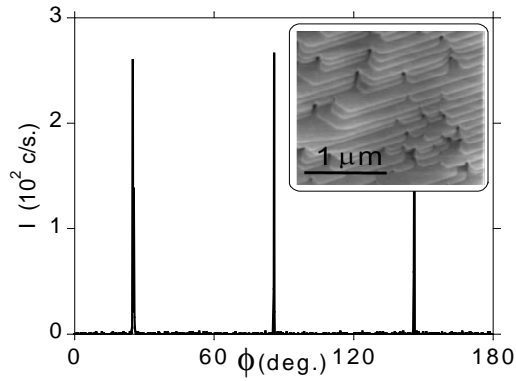

**Figure 2**. X-ray (CuK$_{\alpha 1}$, $\phi$ -) scan of the (300 nm) Bi$_2$Te$_3$ film grown at the mica substrate. According to en expectations the scan peaked after every 60 degrees. AFM image of the film free surface is shown in the insert. One nanometer height growth steps are clear resolved at the image.

Density of the stable nuclei at the mica surface was in the range $10^4$-$10^5$ cm$^2$. Thickness of the Bi$_{2-x}$Sb$_x$Te$_3$ nuclei before coalescence was dependent on x in chemical formula and on $T_S$. Growth steps of a 1 nm height were clear resolved on AFM images of a free surface of the grown thermoelectric films, see insert in figure 2.

C-axis of the thermoelectric film was normal to substrate plane, the (Bi,Sb)$_2$Te$_3$ layers were well in-plane preferentially oriented as well, as follows from the X-ray data show in figures 1 and 2. The only peaks observed in the $\omega/2\theta$ X-ray scans measured for the grown films are (000.n). So the grown thermoelectric layers were free from inclusions of the second phases. The c- and a-axis parameters of unit cell of the grown (Bi,Sb)$_2$Te$_3$ films are 30.46±0.01Å and 4.39±0.01Å, respectively.

## 3. Thermoelectric properties

The temperature dependences of the Seebeck coefficient ($\alpha$) and electroconductivity ($\sigma$) of the Bi$_2$Te$_3$ and Bi$_{0.5}$Sb$_{1.5}$Te$_3$ films are shown in figure 3.

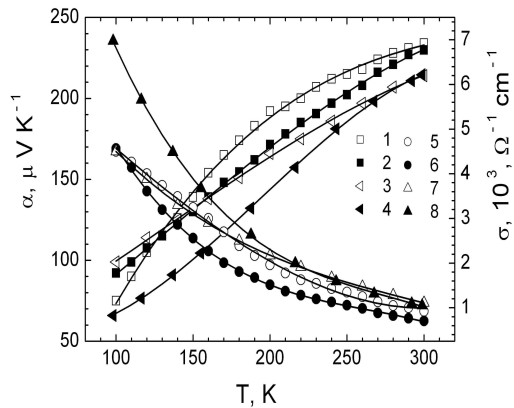

**Figure 3.** Temperature dependences of the Seebeck coefficient $\alpha$ (1-4), electroconductivity $\sigma$ (5-8) for Bi$_2$Te$_3$ (1, 5) and Bi$_{0.5}$Sb$_{1.5}$Te$_3$ (3, 7) films, and for Bi$_2$Te$_3$ (2, 6) and Bi$_{2-x}$Sb$_x$Te$_3$ (4, 8) bulk samples.

Enhancement of Seebeck coefficient and change of slope of its temperature dependence indicate to changes of charge carrier scattering mechanisms and the parameters of the constant-energy surface in grown films [10, 11] in comparison with those in bulk thermoelectric materials (Fig 3, curves 1, 2 and 3, 4, Fig 4, curves 1, 3).

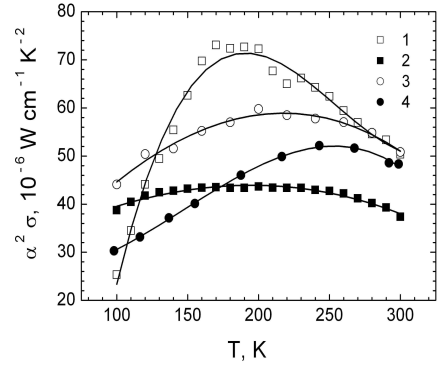

**Figure 4.** Temperature dependences of the power factor $\alpha^2\sigma$ for $Bi_2Te_3$ (1) and $Bi_{0.5}Sb_{1.5}Te_3$ (3) films, and for $Bi_2Te_3$ (2) and $Bi_{2-x}Sb_xTe_3$ (4) bulk samples.

The large values of the power factor together with decrease in thermal conductivity of the films due to selective phonon scattering lead to growth of the figure of merit Z [4-6].

**4. Galvanomagnetic properties and mechanisms of charge carriers scattering**

The data about changes of charge carriers scattering mechanisms in films and bulk thermoelectrics can be obtained from comparative analysis of the galvanomagnetic and thermoelectric properties. The galvanomagnetic properties were studied for epitaxial $Bi_{2-x}Sb_xTe_3$ (x=1.5) films in a wide range of magnetic field to 14 T in the temperature range 77-300 K.

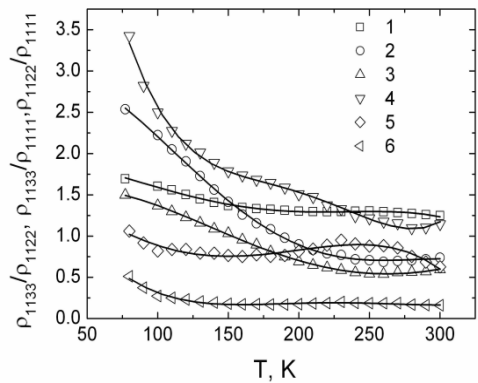

**Figure 5.** Temperature dependence of the ratios of magnetoresistivity tensor components in $Bi_2Te_3$ (1-3) and $Bi_{0.5}Sb_{1.5}Te_3$ (4-6) films.
$\rho_{1133}/\rho_{1122}$ – 1, 4, $\rho_{1133}/\rho_{11111}$ – 2, 5, $\rho_{1122}/\rho_{11111}$ – 3, 6.

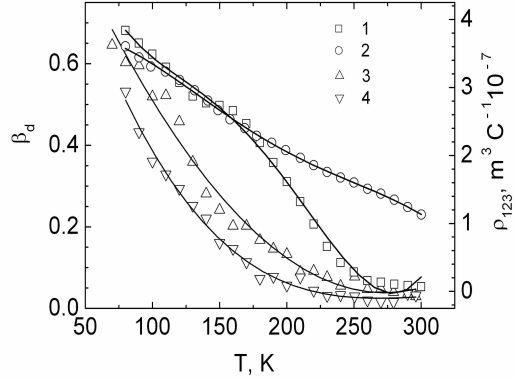

**Figure 6.** Temperature dependence of the Hall effect tensor component $\rho_{123}$ (1, 2) and the degeneracy parameter $\beta_d$ (3, 4) in $Bi_2Te_3$ (1, 3) and $Bi_{0.5}Sb_{1.5}Te_3$ (2, 4) films.

The magnetoresistivity tensor components measured at different orientations of magnetic and electric fields were analyzed with many valley model of energy spectrum of charge carriers for isotropic scattering mechanism (figures 5 and 6)

The least square analysis of experimental data using the many-valley model of the energy spectrum of charge carriers and isotropic scattering mechanism permits to determine degeneracy parameter ($\beta_d$), and some parameters of the constant energy ellipsoidal surfaces: such as effective masses and orientations of ellipsoids relative to the crystallographic axes that characterized anisotropy of the constant energy surfaces of films [10, 11]. The temperature dependence of band structure parameters of the films is supposed to be explained by an additional charge carrier scattering on interphase and interctystallite boundaries of monocrystalline grains of the films (figure 6).

The degeneracy parameter $\beta_d$ in the frame work of the many-valley model can be written as [14]:

$$\beta_d = \frac{I_1^2}{I_0 I_2}, \qquad (1)$$

where $I_0$ is the electrical conductivity, $I_1$ is the Hall conductivity, $I_2$ is the magnetoconductivity. The general expression for $I_n$ (n= 1, 2, 3) defined as:

$$I_n = \left(\frac{e}{m}\right)^n \frac{e^2}{3\pi^2 m} \left(\frac{2m}{\pi^2}\right)^{3/2} \frac{1}{|\alpha_{ij}|^{1/2}} \int_0^\infty \tau^{n+1} \varepsilon^{3/2} \frac{\partial f_0}{\partial \varepsilon} d\varepsilon, \qquad (2)$$

where an isotropic relaxation time $\tau = \tau_0 E^r$, and $\tau_0$ is an energy-independent factor.

The parameter $\beta_d$ for isotropic relaxation time is given by [12]:

$$\beta_d(r,\eta) = \frac{(2r+3/2)^2 F_{2r+1/2}^2(\eta)}{(r+3/2)(3r+3/2)F_{r+1/2}(\eta)F_{3r+1/2}(\eta)}, \quad (3)$$

here $F_s(\eta)$ are Fermi integrals of the type:

$$F_s(\eta) = \int_0^\infty \frac{x^s}{e^{x-\eta}+1}dx, \quad (4)$$

and $\eta$ is the reduced Fermi level.

The experimental values of the Seebeck coefficient $\alpha$ depend on Fermi level and scattering parameter in the form:

$$\alpha = \frac{k}{e}\left[\frac{(r+2.5)F_{r+1.5}(\eta)}{(r+1.5)F_{r+0.5}(\eta)} - \eta\right], \quad (5)$$

The Fermi level ($\eta$) and effective scattering parameters ($r_{eff}$) that accounts changing of the scattering mechanism in the films were obtained by solve two equations (3, 4) for degeneracy parameter $\beta_d$ (r, $\eta$) and the Seebeck coefficient $\alpha(\eta, r)$ using Nelder-Mead least square method [12]. Residual factor $R(r_{eff}, \eta)$ [15] that characterized a discrepancy between experimental and calculated values of accepted physical model were evaluated less than $10^{-5}$ that permits to estimate $\eta$ and r with accuracy less than $10^{-2}$ (figures 7 and 8).

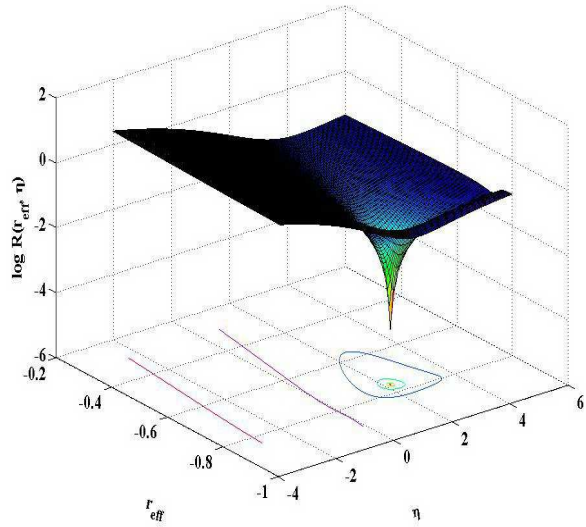

**Figure 7.** Residual factor $R(r_{eff}, \eta)$ calculated for $Bi_2Te_3$ film at 100 K for $\eta$=2.7, r=-0.78.

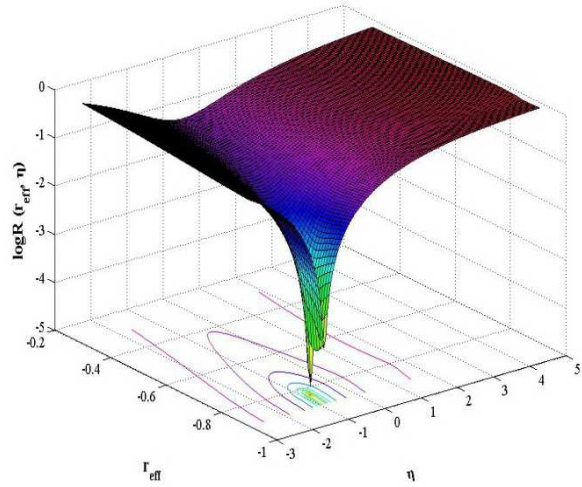

**Figure 8.** Residual factor R($r_{eff}$, η) calculated for $Bi_2Te_3$ film at 300 K for η= -0.87, r= -0.84.

As the result of the experimental data analysis, it was shown that for films and bulk bismuth telluride based materials the value of effective scattering parameter $r_{eff}$ of charge carriers are differed markedly from widely used acoustic scattering parameter r = 0.5 due to influence of charge carriers of the second valence band specific for bismuth telluride and its solid solutions.

The change of the parameter $r_{eff}$ occurs not only due to second valence band of p-$Bi_2Te_3$ known for bulk materials [16] and strong electron-photon interaction but also due to an additional scattering of charge carriers on interphase and interctystallite boundaries of monocrystalline blocks of the films (figure 9).

As follows from figures 10, the dependence of the degeneracy parameter $β_d$ on the reduced Fermi level η shows that the $β_d$ values in films are less, than in bulk materials. Therefore the degeneracy of films is smaller comparing with bulk thermoelectrics with the same compositions (figure 10), since the $β_d$=1 at full degeneration. These specific dependences controlling scattering parameter of charge carriers lead to increase the incline of the Seebeck coefficient dependence on temperature and enhance the thermoelectric power factor for films.

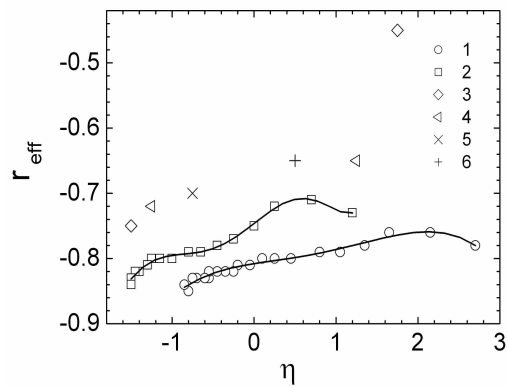

**Figure 9.** The effective scattering parameter $r_{eff}$ for various values of reduced Fermi level η in $Bi_2Te_3$ (1) and in $Bi_{0.5}Sb_{1.5}Te_3$ (2) films.

Similar data for bulk samples of solid solutions: 3, 4 - $Bi_{2-x}Sb_xTe_{3-y}Se_y$ (x=1.2, y=0.09), 5 - $Bi_{2-x}Sb_xTe_{3-y}Se_y$ (x=1.3, y=0.07), 6 - $Bi_{2-x}Sb_xTe_3$ (x=1.6).

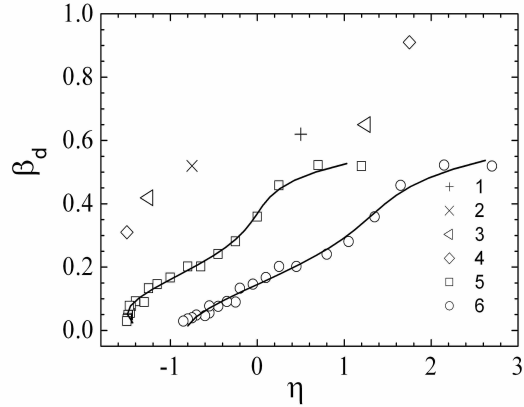

**Figure 10.** The degeneracy parameter $\beta_d$ for various values of reduced Fermi level $\eta$ in $Bi_2Te_3$ (1) and in $Bi_{0.5}Sb_{1.5}Te_3$ (2) films. Similar data for bulk samples of solid solutions: 3, 4- $Bi_{2-x}Sb_xTe_{3-y}Se_y$ [9] (x=1.2, y=0.09), 5 - $Bi_{2-x}Sb_xTe_{3-y}Se_y$ (x=1.3, y=0.07), 6- $Bi_{2-x}Sb_xTe_3$ (x=1.6).

## 5. Conclusions

Epitaxial nanotextured films of $Bi_2Te_3$ and solid solution $Bi_{0.5}Sb_{1.5}Te_3$ were prepared by hot wall technique on freshly cleaved mica (muscovite) substrates. X-ray diffraction studies showed that the *c*-axis of $Bi_2Te_3$ in the films is normal to the substrate plane. Atomic force microscope (Nanoscope IIIa) analysis of film surfaces reveals a system of roughly equidistant growth steps with 1 nm height.

From analysis of the experimental temperature dependences of the Seebeck coefficient and the electrical conductivity of the grown $Bi_2Te_3$ and $Bi_{0.5}Sb_{1.5}Te_3$ films follows that their power factor is higher than the best values for corresponding bulk samples. That may be in part addressed to mechanical stresses affecting charge carrier scattering mechanism.

The charge carrier scattering mechanisms of $Bi_2Te_3$ and solid solution $Bi_{0.5}Sb_{1.5}Te_3$ films were investigated from analysis of the full set of experimental galvanomagnetic coefficients including transverse and longitudinal components of magnetoresistivity tensor, electroresistivity and Hall coefficient with account of the degeneracy parameter $\beta_d$ within many valley model of energy spectrum.

The effective scattering parameter $r_{eff}$ and the reduced Fermi level $\eta$ were calculated by Nelder-Mead least square method from the temperature dependences of the degeneracy parameter $\beta_d$ and the Seebeck coefficient $\alpha$. The parameter $r_{eff}$ in the films is considerably differed from value r=-0.5, specific for an acoustic phonon scattering mechanism as compared to bulk materials. That is explained by an additional charge carriers scattering on interphase and interctystallite boundaries of epitaxial films of chalcogenides of bismuth and antimony. Also the obtained dependence of the degeneracy parameter $\beta_d$ on the reduced Fermi level $\eta$ indicates to weak degeneration in the films.

Revealed charge scattering mechanism peculiarities effect on transport properties of films that enable to optimize thermoelectric parameters and promote to increase thermoelectric power factor and the figure of merit.

**Acknowledgements**

This study was partially supported by Russian Foundation for Basic Research Project No. 13-08-00307a. Staff of the International Laboratory of Low Temperatures and High Magnetic Fields (Wroclaw, Poland) is acknowledgements of the film magnetotransport parameters.